\documentclass[11pt,twoside]{article}
\usepackage{asp2004}
\usepackage{psfig}
\usepackage{epsf}
\usepackage{epsfig}
\usepackage{graphics}
\usepackage{lscape}
\begin{document}

\newcommand\eex[1]{\mbox{$\times 10^{#1}$}}              
\newcommand\eez[1]{\mbox{$10^{#1}$}}                     
\newcommand\nodata{ ~$\cdots$~ }%
\newcommand\nd{\nodata}
\newcommand{\D}{\displaystyle}
\newcommand\hi{\ion{H}{I}}
\newcommand\hii{\ion{H}{II}}
\newcommand\civ{\ion{C}{IV}}
\newcommand\ovi{\ion{O}{VI}}
\newcommand\arcdeg{\mbox{$^\circ$}}%
\newcommand\elec{\mbox{e$^{-}$}}%
\newcommand\fday{\fd}
\newcommand\fhour{\fh}
\newcommand\fmin{\fm}
\newcommand\fsec{\fs}
\newcommand\farcdeg{\fdg} 
\newcommand\farcmin{\farcm}
\newcommand\farcsec{\farcs}
\newcommand\halpha{\mbox{H$\alpha$}}
\newcommand\vdev{\mbox{$v_{\rm dev}$}}
\newcommand\vhel{\mbox{$v_{\rm helio}$}}
\newcommand\msun{\hbox{$M_{\odot}$}}
\newcommand\mass{\msun}
\newcommand\mdot{\hbox{$\dot M$}}
\newcommand\zsun{\mbox{$Z_{\odot}$}}
\newcommand\rsloan{\hbox{$r^\prime$}}
\newcommand\sqarc{\hbox{arcsec$^{2}$}}
\newcommand\psqarc{\hbox{arcsec$^{-2}$}}
\newcommand\lsun{\hbox{$L_{\odot}$}}
\newcommand\SN{signal-to-noise ratio}
\newcommand\kps{\mbox{${\rm km~s^{-1}}$}}
\newcommand\snu{\mbox{$_\nu$}}
\newcommand\tb{\mbox{$T_b$}}
\newcommand\ts{\mbox{$T_s$}}
\newcommand\slambda{\mbox{$S_{\lambda}$}}
\newcommand\cm{\mbox{${\rm cm^{-2}}$}}
\newcommand\nh{\mbox{$N_{\rm HI}$}}
\newcommand\NH{\nh}
\newcommand\mh{\mbox{$M_{\rm HI}$}}
\newcommand\mhi{\mbox{$M_{\rm HI}$}}
\newcommand\mtot{\mbox{$M_{\rm tot}$}}
\renewcommand\mp{\mbox{$m_p$}}
\newcommand\sunits{\mbox{${\rm Jy~beam^{-1}}$}}
\newcommand{\etal}{{\it et al.\/}}
\newcommand{\ie}{{\it i.e.\/}}
\newcommand{\eg}{{\it e.g.\/}}
\newcommand{\pv}{\mbox{$p$-$v$}}
\newcommand{\bc}{\begin{center}}
\newcommand{\ec}{\end{center}}

\title{High-Velocity Clouds in M 83 and M 51}
\author{Eric D. Miller}
\affil{Center for Space Research, Massachusetts Institute of Technology,
Cambridge, MA 02139, USA}
\author{Joel N. Bregman}
\affil{Department of Astronomy, University of Michigan, Ann Arbor, MI
48109, USA}

\begin{abstract}
Various scenarios have been proposed to explain the origin of the Galactic
high-velocity clouds, predicting different distances and implying widely
varying properties for the Galaxy's gaseous halo.  To eliminate the
difficulties of studying the Galactic halo from within, we have embarked on
a program to study anomalous neutral gas in external galaxies, and here we
present the results for two nearby, face-on spiral galaxies, M 83 and M 51.
Significant amounts of anomalous-velocity \hi\ are detected in deep VLA
21-cm observations, including an extended, slowly rotating disk and several
discrete \hi\ clouds.  Our detection algorithm reaches a limiting \hi\
source mass of 7\eex{5} \msun, and it allows for detailed
statistical analysis of the false detection rate.  We use this to place
limits on the HVC mass distributions in these galaxies and the Milky Way;
if the HVC populations are similar, then the Galactocentric HVC distances
must be less than about 25 kpc.  
\end{abstract}

\thispagestyle{plain}

\section{Introduction}

The high-velocity clouds (HVCs) of the Milky Way galaxy were discovered
over 40 years ago by \cite{Mulleretal63}, and have been extensively studied
in \hi\ emission, \halpha\ emission, and absorption, but their nature
remains puzzling.  These clouds of \hi\ gas, deviating from the disk
velocity by up to 300 \kps, exhibit a tremendous variation in size,
morphology and velocity.  The largest complexes span many degrees on the
sky and possess complicated spatial structure, while the compact
high-velocity clouds \citep[CHVCs; ][]{BraunBurton99} are of small angular
extent.  The distances to the HVCs remain largely unknown, except for a few
with direct determinations \citep[\eg, absorption against background halo
stars,][]{Wakker01} or indirect determinations
\citep[\eg,][]{Bland-Hawthornetal98,Putmanetal03}.

A number of formation theories posit either galactic or extragalactic
origins for the HVCs, predicting different physical and dynamical
properties.  Among the galactic models, the most favored is the galactic
fountain, a cyclic phenomenon whereby hot ($T \sim \eez{6}$ K) gas is
ejected by supernova explosions into the lower halo, where it cools and
falls back down as observable \hi\ \citep{ShapiroField76,Bregman80}.  Models
which point to an extragalactic origin include stripping from companion
galaxies, accretion from the inter-galactic medium 
\citep[IGM; \eg,][]{Oort66,Oort70,Oort81}, or a population of gaseous,
dark-matter-dominated mini-halos, either concentrated within $\sim 200$ kpc
of our galaxy or scattered throughout the Local Group
\citep[\eg,][]{Oort66,Blitz99}.  Since the inferred mass of a
21cm-emitting \hi\ cloud goes as the distance squared, the mass
distribution of the HVC ensemble is quite different under the various
scenarios, as summarized in Table \ref{tab:scenarios}.

\begin{table}[t]
\caption{Predictions of HVC Production Scenarios \label{tab:scenarios}}
\smallskip
\begin{center}
\begin{tabular}{lrrrrr}
\tableline
\noalign{\smallskip}
scenario & distance & $|\vdev|$ & size & $M_{\rm cloud}$ & $Z$ \\
 & (kpc) & (\kps) & (kpc) & (\msun) & (\zsun) \\
\noalign{\smallskip}
\tableline
\noalign{\smallskip}
IGM infall          & $< 3$        & $< 150$  & $< 1$     & $<$\eez{4} 
 & 0.1--0.3 \\
galactic fountain   & $< 10$       & $< 150$  & $< 1$  & \eez{4}--\eez{5} 
 & $\ge$ 1 \\
companion stripping & 5--100       & $< 300$  & $\sim 1$  & \eez{5}--\eez{6} 
 & 0.1--1 \\
circumgalactic DM halos & $\sim 150 $  & $< 300$  & $\sim 1$ & \eez{6} 
 & 0.1--0.3 \\
Local Group DM halos    & $\sim 750 $  & $< 300$  & 1--10 & \eez{8} 
 & 0.1--0.3 \\
\noalign{\smallskip}
\tableline
\noalign{\smallskip}
\multicolumn{6}{l}{\footnotesize Sizes and masses shown are typical values based on the median observed parameters      } \\
\multicolumn{6}{l}{\footnotesize of the HVC ensemble.  Masses for the first three scenarios assume \hi\ is the primary  } \\
\multicolumn{6}{l}{\footnotesize constituent of the cloud, while the mini-halo scenarios assume an \hi\ to dark matter } \\
\multicolumn{6}{l}{\footnotesize ratio of 0.1.  Velocities for the galactic fountain and IGM infall models assume an } \\
\multicolumn{6}{l}{\footnotesize adiabatic corona of fully ionized gas with $T \sim \eez{6}$ K. } \\
\end{tabular}
\end{center}
\end{table}

The distance and mass distributions provide fundamental demarcations for
the formation scenarios, and along with metallicity they can select one or
another.  Mass depends on distance, however, and HVC distance is difficult
to determine for clouds of primarily neutral hydrogen.  This problem can be
largely overcome by observing external galaxies of known distance and of
similar type to the Milky Way.  Evidence of anomalous \hi\ sources in spiral
galaxies has been observed for over a decade
\citep[\eg][]{vanderHulstSancisi88,Kamphuisetal91,KamphuisBriggs92,KamphuisSancisi93},
and a recent study has revealed for the first time an extensive population
of discrete \hi\ clouds around an external galaxy
\citep[M31;][]{Thilkeretal04}.

This contribution describes the first results of a study of nearby spiral
galaxies to search for HVC counterparts and other anomalous \hi.  Deep VLA
observations of M 83 and M 51 have revealed several discrete \hi\ sources.  
Comparison of the derived source mass distributions to the expected
Galactic HVC distribution allows us to discriminate between the various HVC
production scenarios.  Full details of the analysis are presented by
\citet{Miller03} and in papers currently in preparation.

\section{Sample Selection and Observations}

The project requires observations of large, late-type spiral galaxies
located nearby to increase \hi\ mass sensitivity and spatial resolution.
Orientation effects limit the visibility of anomalous clouds; while edge-on
galaxies allow discrimination of features in the galactic $z$ direction,
nothing can be known about the $z$ velocity.  In contrast, galaxies
viewed face-on reveal vertical velocity structure, but line-of-sight
positions are degenerate, and high-$z$ sources merge with the disk emission
if they are sufficiently low in $z$ velocity.  Inclined galaxies avoid
these effects to some extent, as has been demonstrated with NGC 2403 
\citep[][and elsewhere in these proceedings]{Fraternalietal02a}, although
the parameter space probed is somewhat lessened.  Overcoming these
degeneracies requires a sample covering a range of inclinations.

We have begun this project by observing two bright, nearby, face-on
spiral galaxies, M 83 and M 51.  The \hi\ 21-cm emission line observations
were taken in 1999 with the VLA in D array, with a spatial resolution of
about 40\arcsec, an effective bandwidth of $\pm 270$ \kps\ (centered on
the systemic velocity), and a velocity resolution of 5.2 \kps\ after
processing.  The primary beam HPBW of 30\arcmin\ provided an effective
field of view of diameter 40 kpc for M 83 (4.5 Mpc distant) and 75 kpc for
M 51 (8.4 Mpc distant).  The dirty image cubes, with effective exposure
times of about 12 hr, were deconvolved with the Multi-Resolution Clean of
\citet{WakkerSchwarz88}. Deep optical observations were taken with the
Curtis Schmidt telescope (M 83, in 1999) and MDM 1.3m telescope (M 51, in
2002) to search for stellar streams associated with anomalous \hi.

\section{Large-Scale Anomalous \hi}

\begin{figure}[t]
\centering{\epsfig{file=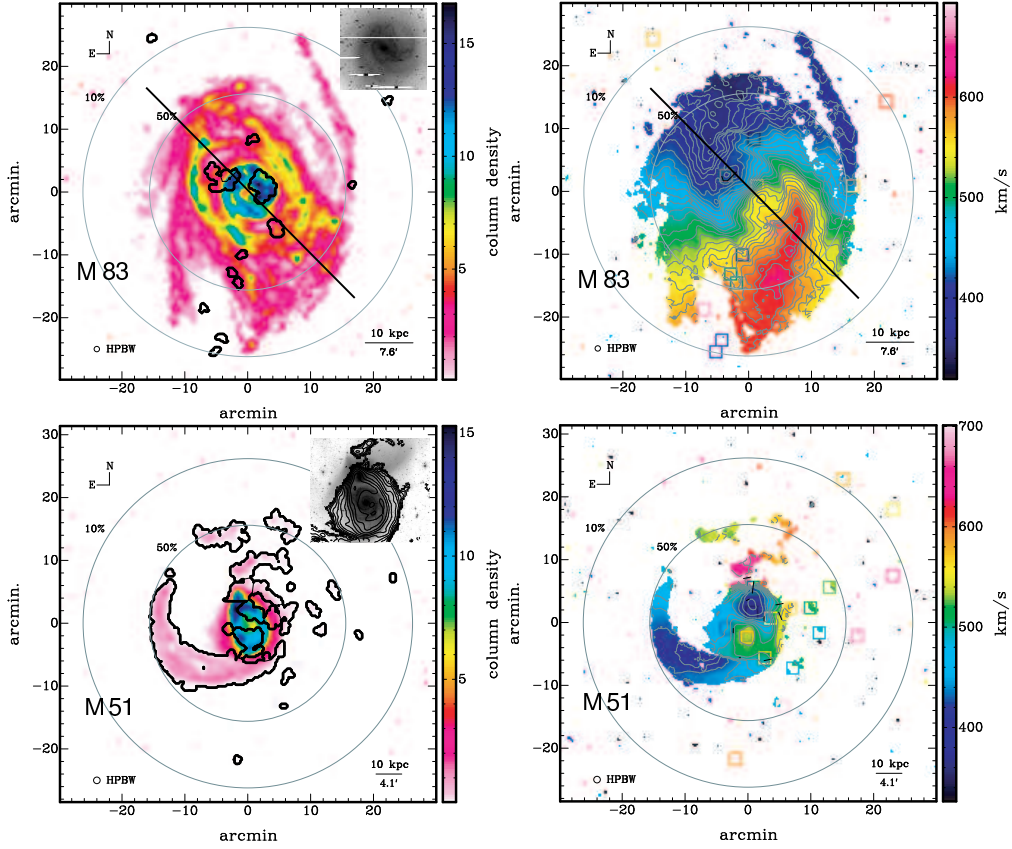,width=\linewidth}}
\caption
{\hi\ column density (in \eez{20} \cm\ units) and mean velocity maps for M
83 and M 51.  The circles indicate the half-power and one-tenth-power
widths for the primary beam.  The black contours in the column density maps
show the locations of the detected AVCs, while the color of the squares in
the velocity maps show their velocities. Optical images are shown at the
same scale.  The diagonal line in the M 83 maps shows the slice taken in
Figure \ref{fig:beard}.}
\label{fig:moments}
\end{figure}

The inner regions of both galaxies exhibit \hi\ structure and kinematics
typical of spiral galaxies, with a central hole in the \hi\ distribution
and a ``spider diagram'' velocity map (see Figure \ref{fig:moments}).  Both
galaxies also contain \hi\ well outside of the optically-defined disk, with
80\% of the \hi\ in M 83 found here \citep{HuchtBohn81} and with M 51
exhibiting long arcs of \hi\ emission unassociated with starlight
\citep[see also][]{Rotsetal90}.  While these are extreme cases, spiral
galaxies often possess extended disks of \hi.

\begin{figure}[p]
\begin{minipage}[t]{\linewidth}
\centering{\epsfig{file=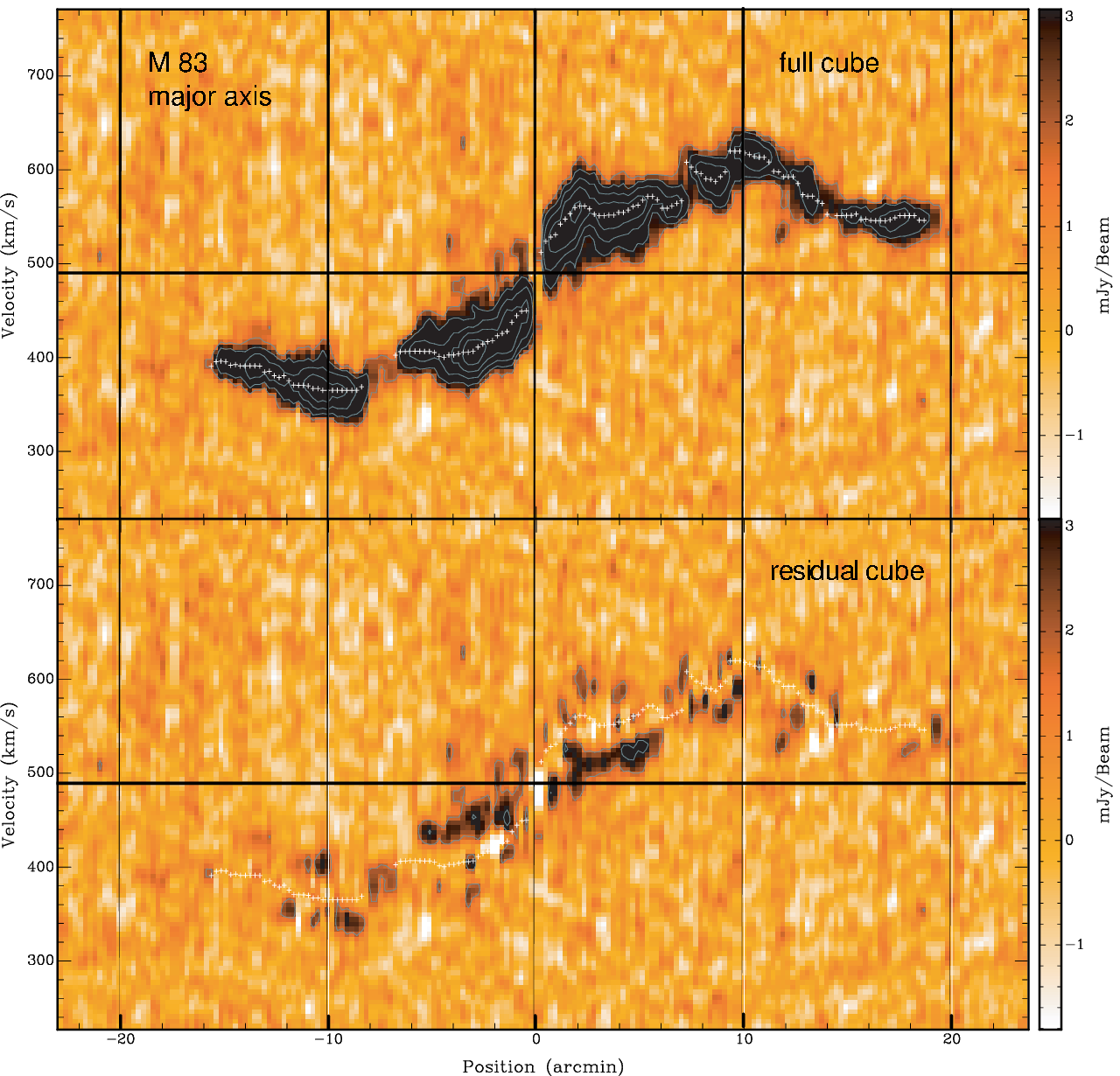,width=.9\linewidth}}
\end{minipage}
\hfill
\begin{minipage}[b]{.45\linewidth}
\centering{\epsfig{file=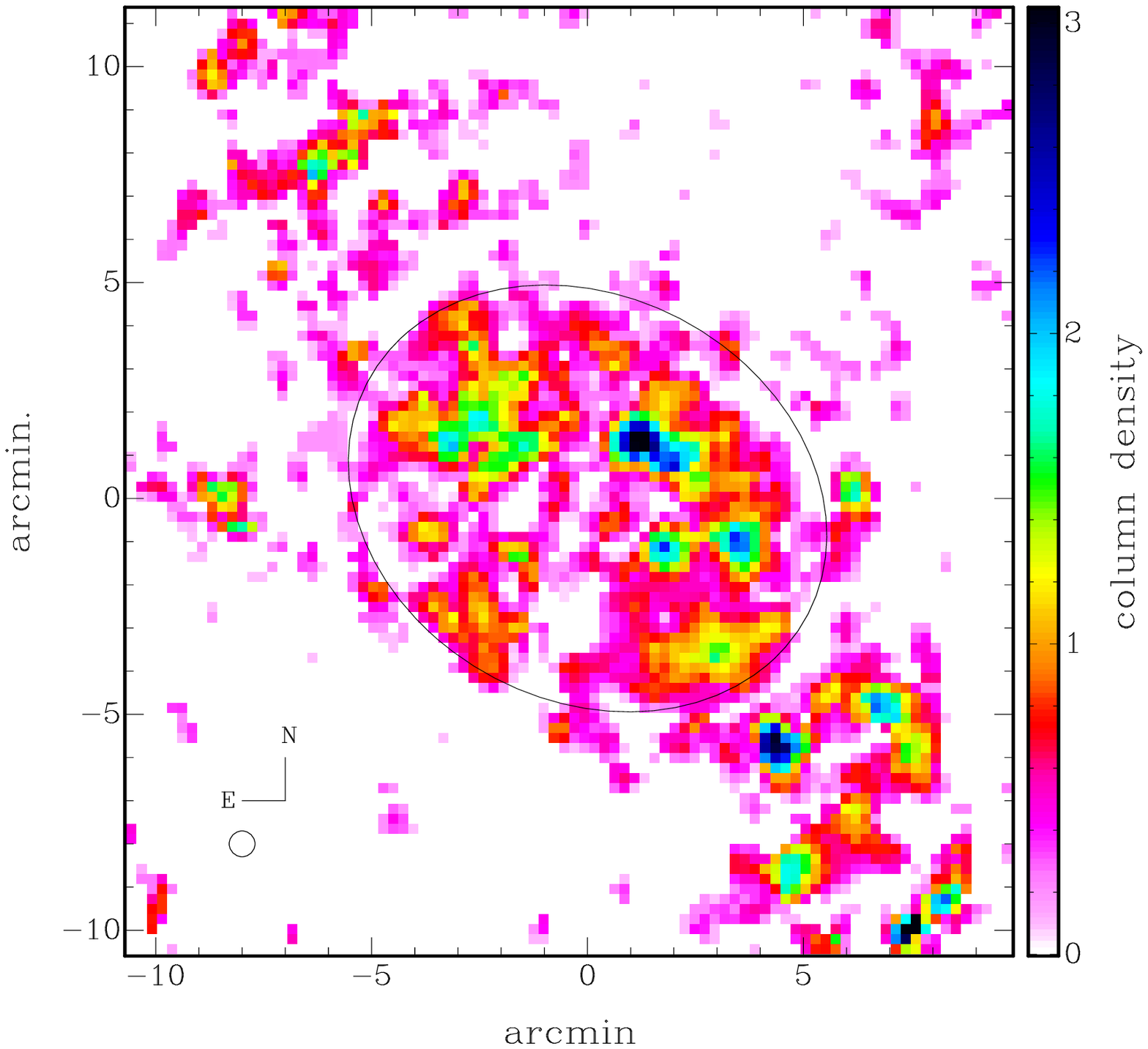,width=\linewidth}}
\end{minipage}
\hfill
\begin{minipage}[b]{.45\linewidth}
\centering{\epsfig{file=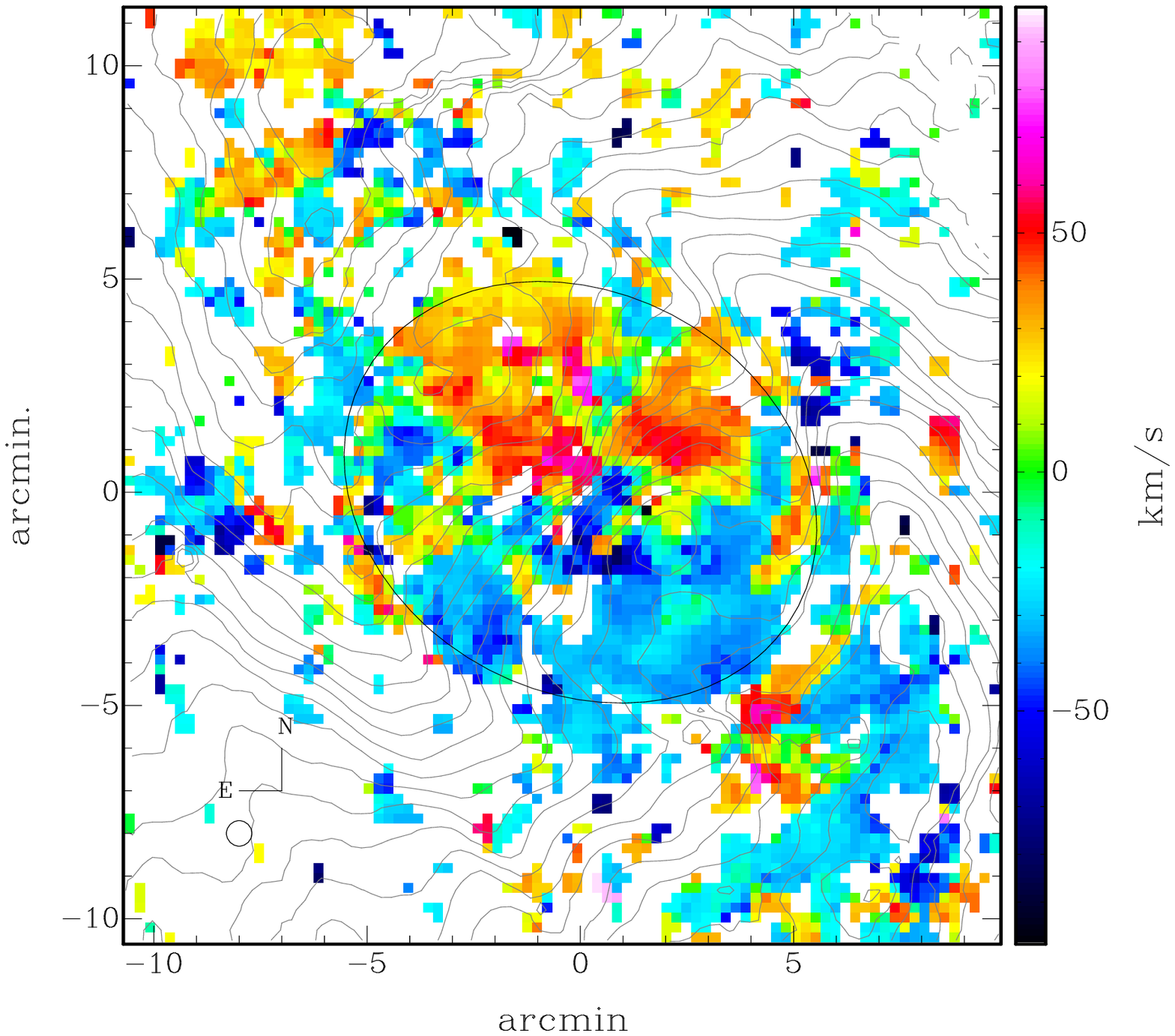,width=\linewidth}}
\end{minipage}
\caption
{{\it Top:\/} Position-velocity plots for the full M 83 \hi\ cube and the
disk-subtracted cube.  The residual cube contains significant emission
between systemic and rotational velocity, which is traced by white
crosses.
{\it Bottom:\/} \hi\ column density (in \eez{20} \cm\ units) and deviation
velocity maps for the disk-subtracted cube of M 83.  Disk-like \hi\
emission remains, and it appears to be rotating 40--50 \kps\ more slowly
the the thin disk in projection.  The contours trace the thin \hi\ disk
velocity, in spacings of 10 \kps, to show its orientation and extent.  The
ellipse traces the inner stellar disk.
\label{fig:beard}}
\end{figure}

Removal of the thin, kinematically cold \hi\ disk is helpful in searching
for anomalous material.  We did this by fitting a Gaussian to the peak of
the velocity profile along every line-of-sight in the data cube and
subtracting this model.  The residual cube contained \hi\ projected on the
disk but at much different velocities, as well as \hi\ emission found in
the wings of the disk profile.  A position-velocity (\pv) cut for the full
M 83 data cube and the residual cube is shown in Figure \ref{fig:beard}
with the slice taken along the major axis (as shown in Figure
\ref{fig:moments}).  Measurable \hi\ is found in the residual cube at
velocities between systemic and the rotation velocity.  The total intensity
map of the residual cube shows that this material is structured into a
clumpy disk, and a map of the difference between the mean disk velocity and
the mean residual \hi\ velocity shows that it is rotating in the same sense
as the thin disk, but 40--50 \kps\ more slowly in projection (see Figure
\ref{fig:beard}).  If the rotation axes of the two components are aligned,
then the anomalous \hi\ is rotating at about 100 \kps, or half the speed of
the thin disk.  The total \hi\ mass of this extended feature is 6\eex{7}
\msun, about 1\% of the total \hi\ mass of the galaxy.  M 51 shows a
similar anomalous disk containing about 3\% of the total \hi\ mass.

The slowly rotating anomalous \hi\ disk is similar to features seen in NGC
891 and NGC 2403.  In the former, \citet{SwatersSancisivanderHulst97} found
a vertically extended layer of \hi\ rotating 25--100 \kps\ more slowly than
the thin disk.  In NGC 2403, \citet{Fraternalietal02a} identified the
``beard'', a layer of \hi\ thought to be rotating 20--50 \kps\ more slowly
than the thin disk.

\section{Small-Scale Anomalous \hi}

Radio astronomy suffers from a lack of robust, statistical, 3-d source
detection software.  In response to this, we developed a suite of
algorithms to search a data cube of semi-independent pixels (as is the case
for radio synthesis imaging), borrowing heavily from the methods developed
by \citet{Uson91} and \citet{WilliamsdeGeusBlitz94}.  The software applies
a matched velocity filter to the data cube, enhancing velocity features
with widths similar to the filtering kernel (a Gaussian).  Statistically
significant peaks are identified and delineated into sources by a 3-d
contouring scheme, down to some specified threshold.  The bright \hi\ disk
is subtracted and/or masked to prevent detection of the profile wings.

Using this technique, we discovered 14 anomalous-velocity clumps (AVCs) in
M 83 and 23 AVCs in M 51.  These discrete emission sources, shown in Figure
\ref{fig:moments}, range from being unresolved to spanning several
synthesized beams.  Some are projected over the stellar disk (typically the
extended ones), while some are found in regions free of starlight down to
27 R magnitudes per square arcsec (typically the compact ones, although M
51 has several large \hi\ features at large radii).  The \hi\ masses are
calculated from the 21-cm luminosities, and these range from 5\eex{5} to
1\eex{8} \msun\ with most in the \eez{6} \msun\ range.  The deviation
velocities range from 50 \kps\ (the lowest we can detect) to 200 \kps.

Some of these detections are likely spurious, and we can use the detection
algorithm to determine the completeness and false discovery rate with
simulated datasets.  Using noise and data cube characteristics similar to
the real data, we simulated 9000 sources of nine different \hi\ masses and
processed them with the detection software.  All sources were detected down
to a mass of 6\eex{5} \msun\ assuming a distance for M 83 (2\eex{6}
\msun\ for M 51).  We constructed 400 data cubes of Gaussian noise and
processed them with the software to obtain the false discovery rate.
Several spurious sources were detected below 1\eex{6} \msun, with
detections above 1.5\eex{6} \msun\ being real HVCs at the 95\% confidence
level for M 83 (5\eex{6} for M 51).  Three of the 14 AVCs in M 83 and 15 of
23 AVCs in M 51 are above this threshold and likely to be real.  The mass
distributions of the low-mass AVCs were compared to the mass distribution
of false detections with a K-S test. The M 83 AVCs were found to be
consistent with spurious detections (K-S probability of 61\%), while the M
51 AVCs were inconsistent with purely false detections, to the 99\%
confidence level (K-S probability of 1\%).  While nothing can be concluded
about the verity of the M 83 detections, there is a statistically
significant excess of low-mass sources in M 51 compared to what is expected
by chance.

\begin{figure}[t]
\centering{\epsfig{file=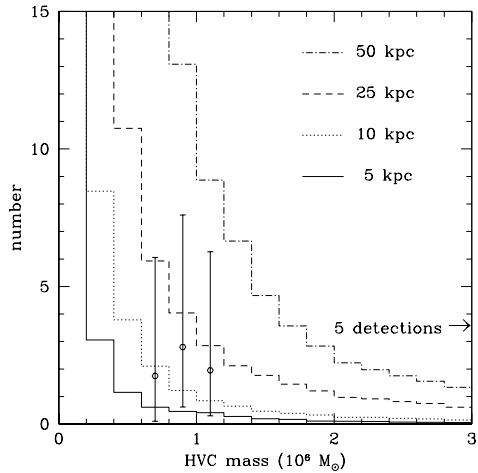,width=.5\linewidth}}
\caption
{Mass distribution of the Milky Way HVCs, assuming uniform spherical
distributions of varying mean distance and using the HVC catalog of
\citet{WvW91}.  The points show the false-detection-corrected distribution
of our lowest-mass detections in M 83, with 95\% confidence errorbars.  If
the HVC mass distributions of the two galaxies are similar, the Milky Way
HVCs must be closer than about 25 kpc or we would have detected more HVCs
in M 83.
\label{fig:massdist}}
\end{figure}

Knowledge of the HVC mass distribution in other galaxies allows us to
constrain the mass distribution (and more importantly the spatial
distribution) of the HVCs observed around our own Galaxy.  We must assume
the HVC population of the Milky Way is similar to these other galaxies in
mass, space, and kinematics.  In addition, we must assume a spatial
distribution for the Galactic HVCs, and here we use a uniform spherical
distribution for simplicity.  Using fluxes tabulated in the all-sky HVC
catalog of \citet{WvW91}, we construct expected mass distributions for a
number of average HVC distances $\langle R \rangle$ by randomly assigning
distances to each AVC.  Four such distributions are shown in Figure
\ref{fig:massdist} along with the mass distribution for the M 83 HVCs,
corrected for false detections.  While the errors are large, the M 83 mass
distribution is consistent with that of the Milky Way HVCs only if the
average HVC distance is less than 25 kpc.  If the Milky Way HVCs are much
further away than that, and if the population is similar to that in M 83,
they would be generally more massive and we would have detected significant
numbers in M 83.

\section{Discussion and Interpretation}

Discovery of a ``beard'' and limits on the HVC mass distribution in M 83
and M 51 constrain the formation models, but as one might expect, multiple
scenarios remain viable, and perhaps required.  Under a galactic fountain,
the slowly rotating and (presumably) vertically-extended \hi\ disk would
represent condensing material beginning to fall back on to the star-forming
disk.  The lagging rotation is in line with expectations of the model, as
material ejected from the disk would flow outward in radius and slow in
rotation.  A possible radial inflow discovered by \citet{Fraternalietal02a}
in NGC 2403 supports this condensing argument as well.  In addition, a
simple estimate of the mass exchange rate from an assumed cooling time is
$\mdot \sim 1$ \msun yr$^{-1}$, a value similar to the star formation rates
of the galaxies under study.

The discrete \hi\ sources are consistent with a galactic fountain.  The
on-disk HVCs, which are real sources of mass $> \eez{6}$ \msun\ in both
galaxies, could be condensed, free-falling clouds or cool gas entrained in
superbubbles breaking out of the galaxy.  The appearance of holes in the
cold \hi\ disk at some of these locations \citep{Miller03} supports the
latter idea, unless the infalling clouds are sufficiently dense to punch
through the disk.

Tidal stripping clearly is active in M 51, with large tidal tails of \hi\
and streams of stars.  Some of the HVC candidates to the west of M 51 might
result from this as well, as it is difficult to reconcile these with a
galactic fountain.  The extended disk and arms of M 83 could be relics of a
recent encounter, of which there is evidence in the form of a faint,
arc-like optical companion \citep{KarachentsevaKarachentsev98}. 

The other extragalactic scenarios remain viable, although this study has
placed constraints on them.  The anomalous \hi\ disk could be accretion
from the IGM; metallicity estimates would be invaluable to determine this.
The discrete AVCs could be dark matter mini-halos, although we detect only
a handful at the $M_{\rm HI} \sim \eez{6}$ \msun\ level.  If they are
about 50 kpc from the Galaxy,  we would have seen nearly five times as many
detections in M 83 and M 51.  Much farther than this, however, and many
would have fallen outside of our field of view.

One thing made clear by this meeting is that most galaxies are found to
contain ``anomalous'' gas upon deep observation.  Warps, extended disks,
and extra-planar clouds are typical of spiral galaxies, and it is the
completely quiescent and regular systems that are the anomalies.

\acknowledgments{We wish to thank the meeting organizers, and especially
Robert Braun, for engineering such an exceptional conference.  Data for
this project were obtained with the VLA, operated by the NRAO, a facility
of the National Science Foundation operated under cooperative agreement by
Associated Universities, Inc.}





\end{document}